\newcommand{\ud}{\rm d}

\documentclass[aps,prl,reprint]{revtex4-1}

\usepackage[T1]{fontenc}
\usepackage{color}
\usepackage{psfrag}
\usepackage{dcolumn}
\usepackage{bm}
\usepackage{amssymb} 
\usepackage[pdftex]{graphicx}
\usepackage{epstopdf}

\begin{document}
\title{Crackling vs. continuum-like dynamics in brittle failure}

\author{J. Barés}
\affiliation{CEA, IRAMIS, SPCSI, Grp. Complex Systems $\&$ Fracture, F-91191 Gif sur Yvette, France}
\author{L. Barbier}
\affiliation{CEA, IRAMIS, SPCSI, Grp. Complex Systems $\&$ Fracture, F-91191 Gif sur Yvette, France}
\author{D. Bonamy}
\affiliation{CEA, IRAMIS, SPCSI, Grp. Complex Systems $\&$ Fracture, F-91191 Gif sur Yvette, France}


\begin{abstract}
We study how the loading rate, specimen geometry and microstructural texture select the dynamics of a crack moving through an heterogeneous elastic material in the quasi-static approximation. We find a transition, fully controlled by two dimensionless variables, between dynamics ruled by continuum fracture mechanics and crackling dynamics. Selection of the latter by the loading, microstructure and specimen parameters is formulated in terms of scaling laws on the power spectrum of crack velocity. This analysis defines the experimental conditions required to observe crackling in fracture. Beyond failure problems, the results extend to a variety of situations described by models of the same universality class, e.g. the dynamics in wetting or of domain walls in amorphous ferromagnets.   
\end{abstract}

\maketitle

Many systems including ferromagnets \cite{Papanikolaou11_natphys}, plastically deformed metals \cite{Richeton05_natmat}, fault seismicity \cite{Rundle03_rg}, liquid spreading \cite{Planet09_prl}, and fracturing solids \cite{Bonamy09_jpd,Rosti09_jpd}  crackle, i.e. respond to a slowly varying external parameter through jerky dynamics, with discrete pulses or avalanches spanning a variety of scales. The salient feature of such crackling dynamics is to exhibit universal scale-free statistics and scaling laws, independent of both microscopic details and external conditions (see \cite{Sethna01_nature} for a review). Those are set by generic properties such as symmetries, dimensions, and interaction range. This behavior, reminiscent of self-organized criticality, is generally explained as being due to the presence of a critical point and a mechanism attracting the system towards this point \cite{Dickman00_bjp}.

In brittle failure problems, the crack front can be identified with a long-range elastic spring \cite{Gao89_jam,Schmittbuhl95_prl,Ramanathan97_prl,Bonamy06_prl}, and the crack onset in heterogeneous/amorphous solids can be mapped to a critical depinnning transition \cite{Roux03_ejma,Ponson09_prl,Bonamy11_pr}. In stable crack growth experiments, crackling dynamics are sometimes observed \cite{Maloy06_prl} and can be attributed to a self-adjustment of the driving force around its depinning value \cite{Bonamy08_prl}. This model is found to reproduce the scaling laws and scale free statistics observed experimentally in \cite{Maloy06_prl}. Still, many situations involving a variety of disordered brittle solids (structural glasses, brittle polymers, ceramics,...) do not exhibit crackling. Rather, they exhibit continuous dynamics compatible with the Linear Elastic Fracture Mechanics (LEFM) predictions. 

By investigating theoretically and numerically crack propagation in elastic disordered media, we reveal that either LEFM-like or crackling dynamics can be observed -- A transition line is exhibited between the two regimes, and defines a phase diagram within a space defined by two reduced variables that intimately mingle the specimen thickness, specimen geometry, loading rate, material constants (fracture energy and crack front mobility), and microstructural texture (disorder contrast and length scale). Within the crackling phase, the Fourier spectrum of the crack velocity is characterized by a power-law with a universal exponent. Conversely the prefactor and the two cutoffs associated to this power-law are found to depend on the loading, microstructure and specimen parameters according to scaling laws that are uncovered herein. These results are discussed within the framework of the depinning theory \cite{Chauve01_prl}. They shed light on the experimental conditions required to observe crackling in brittle fracture. Beyond crack growth problems, they can be immediately extended to a number of others systems described by the same long-range string model, such as the dynamics of contact lines in wetting \cite{Ertas94_pre}, or that of magnetic domain walls with dipolar interactions \cite{Durin00_prl}.

{\em Theory --}  In brittle failure problems, crack destabilization and further propagation are governed by the balance between the amount of elastic energy, $G$, released by the solid as the crack propagates over a unit length, and the fracture energy, $\Gamma$, dissipated in the fracture process zone to create two new fracture surfaces of unit area \cite{Freund90_book}. In standard continuum fracture theory, $G$ depends on the imposed loading and specimen geometry, and $\Gamma$ is a material constant. In the slow fracture regime, the crack velocity $v$ is given by $v/\mu = G-\Gamma$ where (in a perfectly linear elastic material and in the absence of any environmental effect) the effective mobility $\mu$ can be related to the Rayleigh wave speed $c_R$ through $\mu=c_R/\Gamma$.

Defects and inhomogeneities at the microstructure scale yield fluctuations in the local fracture energy: $\Gamma(x,y,z)=\overline{\Gamma}+\gamma(x,y,z)$ where $\hat{x}$, $\hat{y}$, and $\hat{z}$-axis are aligned with the direction of crack propagation, tensile loading, and mean crack front, respectively. This induces $(\hat{x},\hat{z})$ in-plane and $(\hat{y},\hat{z})$ out-of-plane distortions of the front which, in turn, generate local variations in $G$. To the first order, variations of $G$ depend on the in-plane front distortion only. Thus, the problem reduces to that of a planar crack \cite{Ball95_ijf,Movchan98_ijss}. One can then use Rice's analysis \cite{Rice85_jam} to relates the local value $G(z,t)$ of energy release to the planar  front shape $f(z,t)$ (see \cite{Lazarus11_jmps} for a recent review). Once injected in the equation of motion, this yields \cite{Ponson10_ijf}: 

\begin{equation}
\frac{1}{\mu}\frac{\partial f}{\partial t} = F(\overline{f},t) - \overline{\Gamma}J(z,\{f\}) +\gamma(z,x=f(z,t))
\label{EqMotion}
\end{equation}

\noindent where the long-range kernel $J$ is more conveniently defined by its $z$-Fourier transform $\hat{J}(q)=-|q|\hat{f}$. Here, $F(\overline{f},t)=G(\overline{f},t)-\overline{\Gamma}$ and $G(\overline{f},t)$ denotes the mechanical energy release which would result from the same loading conditions with a straight crack front at the mean position $\overline{f}(t)=\langle f(z,t)\rangle_z$. This equation is that of a long-range elastic line driven by this force $F$ within the frozen random potential $\gamma(z,x)$. It exhibits a depinning transition at a critical value $F_c$, characterized by avalanche dynamics and universal scale-free behaviors \cite{Ertas94_pre}. 

The function $G(\overline{f},t)$ is selected by the specimen geometry and imposed loading. It has to be determined using LEFM. In stable growth situations, it should increase with $t$ [crack loaded by imposing external displacements that grow with $t$] and decreases with $\overline{f}$ [specimen compliance increases with $f$]. Without loss of generality, we consider an immobile crack at $t=0$ and we set the $x$-axis origin at its tip ($\overline{f}(t=0)=0$). Then, one gets: $G(\overline{f}=0,t=0)=\overline{\Gamma}$. Considering the subsequent variations $f(z,t)$ are small with respect to the initial crack length, one can write:    

\begin{equation}
F(\overline{f},t) = \dot{G} t - G' \overline{f},
\label{forceVariation}
\end{equation}

\noindent where $\dot{G}=\partial G /\partial t$ (driving rate) and $G'=-\partial G /\partial \overline{f}$ (unloading factor) are positive constants set by the imposed displacement rate and the specimen geometry, respectively.

To complete the description, one has finally to precise the random term $\gamma$ in Eq. \ref{EqMotion}. {\em A priori}, this latter is characterized by the probability function $p(\gamma)$ and the spatial correlation function $C(\vec{r})=\langle \gamma(\vec{r_0}+\vec{r})\gamma(\vec{r_0}) \rangle_{r_0}$. In the following, we will consider (i) a Gaussian distribution $p$ of standard deviation $\tilde{\gamma}$; and (ii) an isotropic correlation function $C$ that decreases linearly with $|r|$ over a distance $\ell$ (correlation length for the disorder landscape) beyond which $C=0$.  Note that the scaling properties are expected to remain unaffected by changing the shapes $p(\gamma)$ and $C(|r|)$ \cite{Vandembroucq04_pre}. Microstructural disorder is then fully characterized by $\tilde{\gamma}$ and $\ell$.

In this framework, the front dynamics are {\em a priori} set by 7 parameters: $\mu$, $\overline{\Gamma}$, $\dot{G}$, $G'$, $\tilde{\gamma}$, $\ell$ and the system size $L$ (specimen thickness along $z$ axis). By introducing the dimensionless time $t \rightarrow t/(\ell/\mu\overline{\Gamma})$ and length $\{x,z,f\} \rightarrow \{x/\ell,z/\ell,f/\ell\}$, one gets:

\begin{equation}
\frac{\partial f}{\partial t} = ct-k\overline{f} -J(z,\{f\}) +\eta (z,x=f(z,t))
\label{motiondimensionless}
\end{equation}

\noindent  where $c=\dot{G}\ell/\mu\overline{\Gamma}^2$ is the dimensionless driving rate, $k = G'\ell/\overline{\Gamma}$ is the dimensionless unloading factor, and $\eta$ is a Gaussian random term of standard deviation $\sigma=\tilde{\gamma}/\overline{\Gamma}$ and unit spatial correlation length. As a result, the front dynamics are selected by four independent parameters, only: $c$, $k$, $\sigma$, and the scale ratio $N=L/\ell$.

{\em Numerics --} Using a fourth order Runge-Kutta scheme, we solved Eq. \ref{motiondimensionless} for a front $f(z,t)$ propagating in a $N \times p N$ uncorrelated random Gaussian map $\eta(z,x)$ with zero average and $\sigma$ variance ($p$ sets the $(\hat{x},\hat{z})$ aspect ratio). The parameter $c$ was varied from $ 10^{-6}$ [imposed by the time limit of 40 days on a 2 GHz CPU we impose for each simulation] to $10^{-4}$ [to keep a large enough scale separation between the depinned front velocity and the loading rate]. The parameters $k$, $\sigma$, and $N$ were respectively varied from $10^{-8}$ to $1$, $10^{-1}$ to $4$, and $32$ to $2048$. This permits a wide exploration of the phase diagram (8 decades in the relevant units, see Fig. \ref{fig2} and associated text).

\begin{figure}
\begin{center}
\includegraphics[width=\columnwidth]{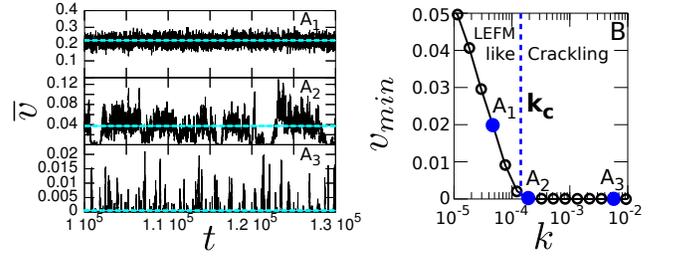}
\caption{A1$\rightarrow$A3: Time evolution of the spatially averaged crack front velocity $\overline{v}(t)$ for increasing unloading factor $k$: $k=4.75 \times 10^{-5}$ (A1), $k=2 \times 10^{-4}$ (A2) and $k=5.5 \times 10^{-3}$ (A3). Other parameters are kept constants: $c=10^{-5}$, $N=1024$ and $\sigma=1$. At low $k$, $\overline{v}(t)$ wanders around the value $c/k$ expected in absence of microstructural disorder, with relative fluctuations that decreases with $k$. For higher $k$, the dynamics become jerky and, above a given value $k_c$, separated pulses can be distinguished, which sharpens as $k$ increases. B: Minimum value of $\overline{v}(t)$ vs. $k$. The transition value $k_c$ between CM-like and crackling dynamics is precisely defined as the smallest value $k$ for which $v_\mathrm{min} =0$.}
\label{fig1}
\end{center}
\end{figure}

\begin{figure}
\begin{center}
\includegraphics[width=\columnwidth]{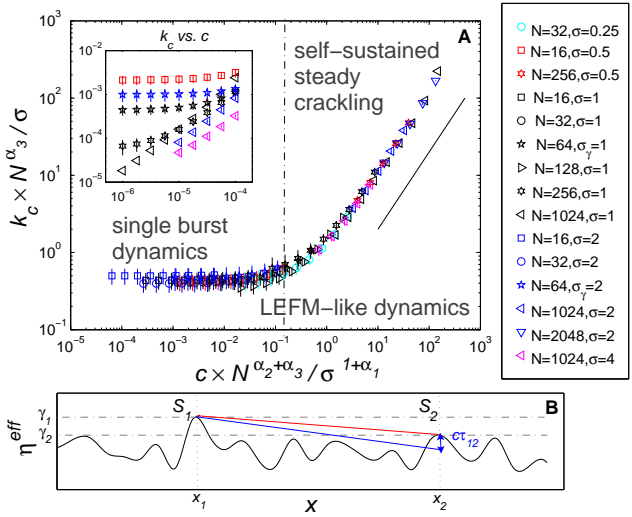}
\caption{(Color online) A: Phase diagram of the crack dynamics. Inset: variation of $k_c$ as a function of $c$ for different $N$ and $\sigma$ (values on right-hand side). Each point results from averaging over many simulations and errorbars correspond to a $95\%$ confidence interval. Main panel: Collapse obtained using Eq. \ref{EqDiagram} with $\alpha_1=0.38$, $\alpha_2=1.15$ and $\alpha_3=1.65$. Straight line indicates proportionality. In both graphs, the axis are logarithmic. B: Sketch of the variation of the effective pinning force applying on the front as it quasistatically propagates throughout the disordered landscape (see e.g. \cite{Tanguy98_pre} for implementation of such a propagation algorithm). Points S$_1=\{x_1,\eta_1\}$ and S$_2=\{x_2,\eta_2\}$ locate the maximum and following next-to-maximum peaks over the traveled distance (see text for details).}
\label{fig2}
\end{center}
\end{figure}

{\em Results --} Figure \ref{fig1}:A1$\rightarrow$A3 presents typical time profiles of the spatially averaged crack velocity $\overline{v}(t)=\ud \overline{f}/\ud t$ for constant $c$ and increasing $k$. At low $k$, $\overline{v}(t)$ fluctuates over the mean value $c/k$ that would have been expected from Continuum Mechanics (CM), i.e. for $\eta = \gamma \equiv 0$. When $k$ increases, the signal becomes more jerky and, above a given value, exhibits crackling dynamics, with distinct pulses separated by silent periods where $\overline{v}=0$. The transition $k_c$ between these two regimes can be computed by plotting the minimum value of $\overline{v}(t)$ as a function of $k$ (Fig. \ref{fig1}:B). $v_{min}$ is equal to zero in the crackling regime, and increases with $k$ in the CM-like regime, above $k_c$. 

On Fig. \ref{fig2}:A(Inset), we plot $k_c \; \mathrm{vs.}\; c$ as measured in systems of fixed $p=3$ and various $N$ and $\sigma$. A $c$-independent plateau ${k_c}_{sat}$ is observed at low $c$/low $N$/large $\sigma$ while $k_c$ increases linearly (slope A) with $c$ at high $c$/high $N$/low $\sigma$. All curves can then be superimposed by making $k_c\rightarrow k_c^* = k_c/{k_c}_{sat}$ and $c \rightarrow c^* = Ac/{k_c}_{sat}$. Both $A$ and ${k_c}_{sat}$ are found to go as a power-law with $N$ and $\sigma$: $A \approx \sigma^{-\alpha_1} N^{\alpha_2}$ with $\alpha_1=1.15\pm0.05$ and $\alpha_2=0.38\pm 0.05$, and ${k_c}_{sat} \approx \sigma N^{-\alpha_3}$ with $\alpha_3=1.65\pm 0.05$. The resulting master function, plotted in Fig. \ref{fig2}:A(main), is:

\begin{equation}
   k^*_c=f(c^*) \quad \mathrm{with} \quad f(c^*)\approx \left\{
\begin{array}{l l}
{k^*_c}_{sat} & \, \mathrm{if} \, c^* \ll {k^*_c}_{sat}  \\
c^* & \, \mathrm{if} \, c^* \gg {k^*_c}_{sat}
\end{array}
\right.
\label{EqDiagram}
\end{equation}

\noindent where $c=c\times N^{\alpha_2+\alpha_3}/\sigma^{1+\alpha_1}$ and $k^*_c=k_c/\sigma N^{-\alpha_3}$. The plateau value ${k^*_c}_{sat}$  is found to decrease with $p$. This curve separates CM-like and crackling dynamics. 

The form of the $k_c \; \mathrm{vs.}\; c$ curves can be understood by analyzing the profile $\eta^{eff}(x)=\langle J(z,\{f\}+\eta(z,x=f(z,t)) \rangle_z$ of the effective pinning force applying on the front as it propagates throughout the disordered landscape. Such a profile is depicted in Fig. \ref{fig2}B. The value ${k_c}_{sat}$ observed for $c \rightarrow 0$ is set by the relative positions of the maximum and the following next-to-maximum peaks over the traveled distance (S$_1=\{x_1,\eta_1\}$ and S$_2=\{x_2,\eta_2\}$ in Fig. \ref{fig2}B): ${k_c}_{sat}=(\eta_1-\eta_2)/(x_2-x_1)$. At finite $c$, the front earns an extra driving force during its depinning jump (duration $\tau_{12}$) from S$_1$ to S$_2$, yielding $k_c={k_c}_{sat}+A c$ with $A=\tau_{12}/(x_2-x_1)$. One thus expects $k_c \approx {k_c}_{sat}$ for $c\ll {k_c}_{sat}/A$ and $k_c \approx A c$ for $c \gg {k_c}_{sat}/A$. The linear variation of $\{\eta^{eff}\}$ with $\sigma$ explains the observed ${k_c}_{sat}\propto \sigma$. Note that, in this scenario, the jerky dynamics observed for $c \ll  {k_c}_{sat}/A$ are dominated by a single large avalanche (from S$_1$ to S$_2$), while true steady self-sustained crackling dynamics can only be observed for $c \gg {k_c}_{sat}/A$.

\begin{figure}
\begin{center}
\includegraphics[width=\columnwidth]{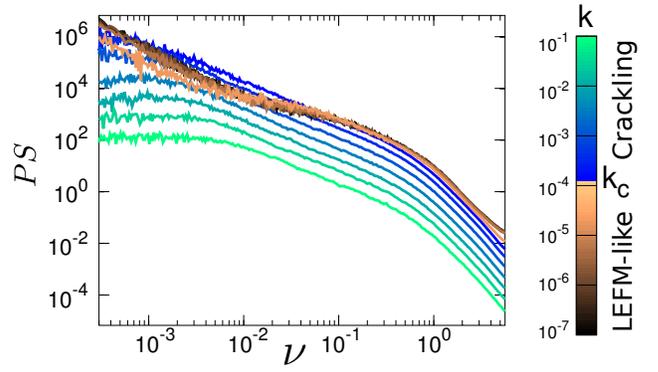}
\caption{(Color online) Power spectrum (PS) of $\overline{v}(t)$ obtained for various value of $k$ (logarithmic axes). Other parameters are kept constant: $c=10^{-5}$, $N=1024$ and $\sigma=1$. Right-handed colorbar indicates the $k$ value. Note the qualitative change at the transition $k_c$, and the power-law observed above, in the crackling regime.}
\label{fig3}
\end{center}
\end{figure}  

We now focus on the evolution of the fracturing dynamics $\overline{v}(t)$ within the steady regimes of the phase diagram. One way to characterize it is to analyze its power spectrum (PS). Such an analysis, indeed, has two advantages with respect to the standard statistical analysis of pulse size and duration developed to analyze crackling signals \cite{Travesset02_prb}: i) it allows a full exploration of the phase diagram (both crackling and LEFM-like); ii) in the crackling part, it does not call for any additional criteria (threshold setting) to filter single pulses in the presence of overlapping avalanches. Figure \ref{fig3} presents the evolution of $PS(\nu)$ for increasing $k$ and  the other parameters constant. Below $k_c$, all curves overlap except at the lowest frequencies. This is precisely what is requested in a CM description, where the continuum-level scale control parameter $k$ should affect the system at large scales only. Conversely, above $k_c$, the PS curves are distinct showing that all scales are affected by $k$. One points out the power-law behavior characteristic of crackling dynamics \cite{Kuntz00_prb,Travesset02_prb,Durin05_book}. The power-law exponent $1/a$ is independent of $k$, whereas the prefactor decreases with $k$. The dramatic change observed as $k$ crosses $k_c$ is a signature that the CM-crackling transition line is a true transition, not a crossover phenomenon. 

\begin{figure}
\begin{center}
\includegraphics[width=\columnwidth]{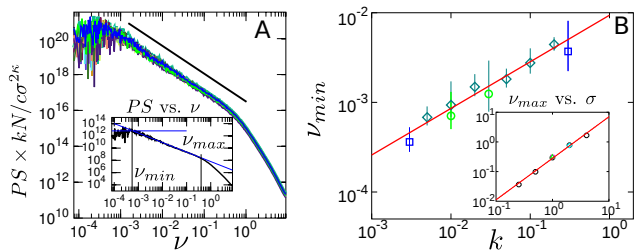}
\caption{(Color online) A(main panel): Collapsed PS curves of $\overline{v}(t)$ in the self-sustained steady crackling part of the phase diagram
(logarithmic axes). The collapse is obtained using Eq. \ref{EqPS} with $\kappa = 0.7$. The different curves (with different colors) correspond to different values of ${c, k, N, \sigma}$: $10^{-6} \leq c \leq 3 \times 10^{-4}$, $3 \times 10^{-3} \leq k \leq 3 \times 10^{-2}$, $128 \leq N \leq 2048$, $\sigma = 1$. Black straight line is a power law of exponent $1/a=1.5$. A(inset): For a given $PS(\nu)$ curve, $\nu_{min}$ is defined from the intersection of the power-law and the plateau value at low frequencies, while $\nu_{max}$ sets the upper cutoff for the power-law. B(main panel): Variation of the lower cutoff $\nu_\mathrm{min}$ with $k$ for $c=10^{-5}$ ($o$), $c=3\times 10^{-5}$ ($\square$), $c=3\times 10^{-5}$ ($\diamondsuit$), and constant values $N=1024$ and $\sigma=1$ [It was checked that $\nu_\mathrm{min}$ is independent of both $N$ and $\sigma$]. In both plots, straight lines are power-law (Eq. \ref{EqCutoff}) with $2\kappa = 1.4$, and $\Delta = 0.52$. B(inset): Variation of the upper cutoff $\nu_\mathrm{max}$ with $\sigma$ for $N=128$ ($\square$), $N=256$ ($\triangle$),$N=1024$ ($o$), $N=2048$  ($\diamondsuit$) and constant values $c=10^{-5}$ and $k=10^{-3}$ [It was checked that $\nu_\mathrm{max}$ is independent of both $c$ and $k$].}
\label{fig4}
\end{center}
\end{figure}  

We turn now to the quantitative selection of the PS in the crackling regime. The curve collapse presented in Fig. \ref{fig4}:A unravels the scaling between the power-law prefactor and the series of variables $c$, $k$,  and $N$: Over the range $\nu_\mathrm{min}\leq \nu \leq \nu_\mathrm{max}$, $PS(\nu)$ is:

\begin{equation}
 PS \approx \frac{\sigma^{2\kappa}}{N} \frac{c}{k}\nu^{-1/a}
\label{EqPS}
\end{equation}
 
\noindent The upper cutoff is found to depend on $\sigma$ only (Fig. \ref{fig4}:B(inset)), while the lower one depends on $k$ only (Fig. \ref{fig4}:B(main)):

\begin{equation}
\nu_\mathrm{max} \approx \sigma^{2\kappa},  \quad \nu_\mathrm{min} \approx k^\Delta 
\label{EqCutoff}
\end{equation}

\noindent In Eqs. \ref{EqPS} and \ref{EqCutoff}, the fitted exponents were found to be $1/a \simeq 1.50 \pm 0.02$, $\kappa \simeq 0.7\pm 0.1$, and $\Delta \simeq 0.52 \pm 0.08$. 

{\em Discussion --} The crackling pulses evidenced in the $\overline{v}(t)$ signal result from the depinning avalanches. Single, non-overlapping, avalanches are known to exhibit universal scale-free distributions and scaling relations characterized by a variety of critical exponents, which can be estimated using renormalization group (RG) \cite{Ertas94_pre,Chauve01_prl} or numerical \cite{Rosso02_pre,Duemmer07_jsm} methods. These scale-free features only hold for length-scales larger than the Larkin length \cite{Larkin79_jltp} $L_c$, which, for our model, scales as $L_c \approx 1/\sigma^2$. We then expect $\nu_\mathrm{max} \approx 1/L_c^\kappa\approx \sigma^{2\kappa}$, where $\kappa=0.770(5)$ \cite{Duemmer07_jsm} refers to the dynamic exponent. This value is in agreement with that measured here. In the so-called adiabatic limit ($c\rightarrow 0$), there is a one to one relation between the $\overline{v}(t)$ pulses and the single depinning avalanches. Then, the PS exponent $a_{ad}$ in Eq. \ref{EqPS} [Here, "ad" index stands for "adiabatic limit"] is expected \cite{Kuntz00_prb} to be the one that defines the scaling $T\propto S^{a_{ad}}$ between the avalanche size $S$ and duration $T$: $a_{ad} = \kappa/(1+\zeta)$ \cite{Bonamy09_jpd} where $\zeta=0.385(5)$ \cite{Rosso02_pre,Duemmer07_jsm} refers to the roughness exponent. As a result, one expects $1/a_{ad}=1.80(2)$. The exponent $\Delta_{ad}$ in Eq. \ref{EqCutoff} defines the scaling between the upper cutoff in time for scale-free features, and the unloading factor $k$. In our model, it is given by $\Delta_{ad} = \kappa/2$ \cite{Bonamy09_jpd}, which yields $\Delta_{ad}=0.385(5)$. Both $\Delta_{ad}$ and $1/a_{ad}$ are found to be significantly different from the values $\Delta$ and $a$ measured herein. By yielding some overlap between the avalanches, a finite driving rate $c$, indeed, is expected \cite{White03_prl} to alter the PS shape and the cutoff dependencies. It is interesting to note that the effect is limited to a novel value set for $a$ and $\Delta$, without modifying the power-law shape for $PS$, nor yielding an additional dependency with $c$ for $\nu_\mathrm{min}$. By yielding percolation throughout the space-time diagram as $c$ increases and/or $k$ decreases, the overlap effect is also believed to drive the crackling/CM transition. On-going work aims at accurately characterizing this coalescence process. This will allow unraveling the selection of $a$ and $\Delta$ in Eq. \ref{EqPS} and that of $\alpha_i$ in Eq. \ref{EqDiagram}.  

To summarize, we have analyzed here how a brittle crack selects its propagation dynamics in the presence of microstructural disorder. Large disorder (contrast or length-scale), large unloading factor, small specimen size and small driving rate yield crackling dynamics, while the opposite yields CM-like dynamics. The associated phase diagram is  unraveled and is shown to be fully controlled by two reduced variables (Fig. \ref{fig2} and Eq. \ref{EqDiagram}) that intimately mingle the above parameters. Relations between these parameters and the dynamics in the crackling phase (Fourier spectrum of the crack velocity) have been finally determined (Eqs. \ref{EqPS} and \ref{EqCutoff}). 

This work sheds light on the experimental conditions required to observe crackling in brittle fracture. It also provides insights on how to decipher the crackling dynamics and gain information on the underlying conditions, e.g. in terms of microstructure or loading when those are not {\em a-priori} known. These results can also inform technological relevant fracture processes, e.g. in the future development of rationalized design methodologies to prevent (or to limit) inopportune crackling (and induced indetermination) in cutting technologies. Beyond solid failure, our analysis directly extends to a number of others systems described by the same long-range string model, such as the dynamics of contact lines in wetting problems \cite{Ertas94_pre} and the dynamics of domain walls in ferromagnets \cite{Durin00_prl} (field sweep rate and demagnetization factor then playing the role of $c$ and $k$). As such, it may be relevant to other fields facing similar problems, e.g. nanofluidic or nanomagnetism technologies.

\begin{acknowledgments}
We thank Alberto Rosso and Alexander Dobrinevski for fruitful discussions, and Cindy Rountree for a critical reading of this manuscript. Support through ANR project MEPHYSTAR is gratefully acknowledged.
\end{acknowledgments}

\end{document}